\documentclass[preprint,12pt,3p]{elsarticle}
\usepackage[utf8]{inputenc}
\usepackage[english]{babel}
\usepackage{hyphenat}
\usepackage{booktabs}
\usepackage{gensymb}
\usepackage{siunitx}
\usepackage{amsmath}
\usepackage{color,soul}
\usepackage[version=4]{mhchem}
\usepackage{subfig}
\usepackage[figuresright]{rotating}
\usepackage{multirow}
\usepackage{setspace}
\usepackage{hyperref}
\usepackage{cleveref}
\usepackage{lipsum}
\usepackage{everysel}
\usepackage[dvipsnames]{xcolor}
\usepackage{wrapfig}
\usepackage{adjustbox}
\biboptions{sort&compress}
\usepackage[cmintegrals,cmbraces]{newtxmath}

\usepackage[sfdefault,condensed]{roboto}  

\begin{document}

\begin{frontmatter}
	
\journal{arXiv}

\title{\raggedright \LARGE\textbf{Transcending the MAX phases concept of nanolaminated early transition metal carbides/nitrides -- the ZIA phases}}

\author[LANL]{M.A. Tunes\corref{cor}}
\cortext[cor]{Corresponding authors:}\ead{m.a.tunes@physics.org}
\author[LANL,UTK]{S.M. Drewry}
\author[LANL,UCB]{F. Schmidt}
\author[MPA]{J.A. Valdez}
\author[LANL]{M.M. Schneider}
\author[LANL]{C.A. Kohnert}
\author[LANL]{T.A. Saleh}
\author[USP]{C.G. Sch\"on}
\author[MPA]{S. Fensin}
\author[PNNL]{O. El-Atwani}
\author[KUL]{N. Goossens}
\author[KUL]{S. Huang}
\author[KUL]{J. Vleugels}
\author[PNNL]{S.A. Maloy}
\author[HUD]{K. Lambrinou\corref{cor}}\ead{k.lambrinou@hud.ac.uk}

\address[LANL]{Materials Science and Technology Division, Los Alamos National Laboratory, United States of America}
\address[UTK]{Department of Materials Science and Engineering, University of Tennessee, United States of America}
\address[UCB]{Department of Nuclear Engineering, University of California Berkeley, United States of America}
\address[MPA]{Materials Physics and Applications Division, Los Alamos National Laboratory, United States of America}
\address[USP]{Department of Metallurgical and Materials Engineering, Universidade de S\~ao Paulo, Brazil}
\address[PNNL]{Reactor Materials and Mechanical Design, Pacific Northwest National Laboratory, United States of America}
\address[KUL]{Department of Materials Engineering, KU Leuven, Belgium}
\address[HUD]{School of Computing and Engineering, University of Huddersfield, United Kingdom}

\begin{abstract}
\doublespacing
\noindent A new potential class of nanolaminated and structurally complex materials, herein conceived as the Zigzag IntermetAllic (ZIA) phases, is proposed. A study of the constituent phases of a specific Nb--Si--Ni intermetallic alloy revealed that its ternary H-phase, \textit{i.e.}, the Nb$_3$SiNi$_2$ intermetallic compound (IMC), is a crystalline solid with the close-packed \textit{fcc} Bravais lattice, the 312 MAX phase stoichiometry and a layered atomic arrangement that may define an entire class of nanolaminated IMCs analogous to the nanolaminated ceramic compounds known today as the MAX phases. The electron microscopy investigation of the Nb$_{3}$SiNi$_{2}$ compound -- the first candidate ZIA phase -- revealed a remarkable structural complexity, as its ordered unit cell is made of 96 atoms. The ZIA phases extend the concept of nanolaminated crystalline solids well beyond the MAX phases family of early transition metal carbides/nitrides, most likely broadening the spectrum of achievable material properties into domains typically not covered by the MAX phases. Furthermore, this work uncovers that both families of nanolaminated crystalline solids, \textit{i.e.}, the herein introduced \textit{fcc} ZIA phases and all known variants of the \textit{hcp} MAX phases, obey the same overarching stoichiometric rule $P_{x+y}A_xN_y$, where $x$ and $y$ are integers ranging from 1 to 6. 

\end{abstract}

\begin{keyword}
MAX Phases, Electron Microscopy, Nanolaminated Crystalline Solids, ZIA Phases
\end{keyword}

\end{frontmatter}

\newpage

\onehalfspacing

\section*{Special Dedication}
\noindent \textit{\small This manuscript is dedicated to V. Nowotny, E. Reiffenstein, F. Benesovsky and W. Jeitschko of the University of Vienna, Austria, who first synthesized complex ternary carbides in the 1960/1970s, and to M.W. Barsoum of Drexel University, USA, who in the 1990/2000s revolutionized this field with his research on the MAX phases and the discovery of their 2D derivatives known as MXenes in 2011. Another dedication goes to E.I. Gladyshevskii, Yu.B. Kuz’ma and P.I. Kripyakevich of the Ivan Franko L’vovsk State University, former USSR, for their pioneering work on complex ternary intermetallics of the Nb$_{3}$Ni$_{2}$Si  type in the 1960/1970s. The herein proposed ``ZIA phases'' are named after the ZIA indigenous nation of New Mexico, USA, where the Los Alamos National Laboratory is located.} 

\section{Introduction} 
\label{sec:intro}

\noindent A revolution in materials science was undoubtedly achieved through the microstructural manipulation of materials on the nanoscale \cite{cahn1990nanostructured,gleiter1992nanostructured}. When the dimensions of the building blocks of a material are confined in the 1-100 nm range, the material can be defined as nanostructured \cite{moriarty2001nanostructured}. The nanoscale manipulation of matter through the engineering of structural modules, such as atomic clusters, precipitates, crystallites/grains, grain boundaries, molecules, and/or atomic layers has been an effective approach to achieve a set of extraordinary and unique properties in nanostructured materials, well beyond what is possible by conventional processing routes that target changes on the micro- and macroscale \cite{cahn1990nanostructured,gleiter1992nanostructured,moriarty2001nanostructured,martin2018rise}. Transferring the superior properties of nanostructured materials to the bulk scale, a process demanding multiscale materials engineering, is one of the key challenges of contemporary materials science \cite{martin2018rise}. 

H. Gleiter – the scientist who introduced the term ``nanostructured materials'' \cite{gleiter1992nanostructured} – predicted in 1992 \cite{gleiter1992nanostructured} that this new research field would be significantly impacted by the emergence of two classes of nanostructured materials: ceramics and intermetallic phases. Gleiter’s prediction was partly realized in 1996 by M.W. Barsoum and his group, who made significant strides in the development and characterization of a class of nanolaminated ceramics with the M$_{n+1}$AX$_{n}$ general stoichiometry, where M is an early transition metal, A is an element mainly from groups 13-15 in the periodic table, X is C or N, and n = 1, 2, or 3; these ceramic materials are known today as the MAX phases \cite{barsoum1996synthesis,barsoum2001max}. The 1996 breakthrough by Barsoum and El-Raghy \cite{barsoum1996synthesis} consisted in the pioneering synthesis of quasi-phase-pure Ti$_{3}$SiC$_{2}$ via reactive hot pressing; notably, the existence of the Ti$_{3}$SiC$_{2}$ ternary carbide, which is the H-phase of the ternary Ti--Si--C system \cite{jeitschko1963kohlenstoffhaltige}, had first been reported in 1967 by Jeitschko and Nowotny \cite{jeitschko1967kristallstruktur}. Starting by the successful synthesis and characterization of Ti$_{3}$SiC$_{2}$, \textit{i.e.}, the 312 ternary carbide of the Ti--Si--C system \cite{jeitschko1963kohlenstoffhaltige}, Barsoum realized early in the 2000s that the Ti$_{3}$SiC$_{2}$ compound and other nanolaminated ternary carbides/nitrides form an entire new class of hexagonal close-packed (\textit{hcp}) ceramics described by the M$_{n+1}$AX$_{n}$ strict stoichiometric rule and characterized by exceptional physicochemical and mechanical properties \cite{barsoum2001max}. To date, around 155 ternary carbides/nitrides of early transition metals as well as numerous chemically complex solid solutions with the M$_{n+1}$AX$_{n}$ strict stoichiometry and precisely tailored properties have been experimentally synthesized \cite{sokol2019chemical,goossens2021microstructure}. The exceptional properties of the MAX phases are attributed to their ordered \textit{hcp} atomic arrangement of M$_6$X ceramic-like octahedra and intercalated A metallic-like layers, which results in nanolaminated unit cells with long-range translational symmetry. Since their inception, the MAX phases \cite{dahlqvist2018origin,li2005participation} and their 2D derivatives known as MXenes \cite{naguib2011two,naguib2012two,gogotsi2019rise,wu2020assembly,long2022roles} have been proposed for various applications in diverse industrial/technological fields, including energy storage, fuel cells, biomedical, photonics/phononics \cite{ghebouli2011theoretical,lapauw2016synthesisA,lapauw2016synthesisB,xu2016demonstration,lin2018insights,tong2019difference,tao2019atomically,naslund2020x,sanna2022fluorinated,shao2023synthesis,zha2021role,saeed20222d,chen2023multiprincipal}, as well as in advanced nuclear fusion and fission reactor systems \cite{tallman2015effect,ang2016anisotropic,tallman2016effect,ang2017microstructure,tallman2017effects,tunes2019transmission,lapauw2019interaction,bowden2020stability,tunca2020compatibility,tunes2021deviating,tunes2022accelerated,tunca2021situ}. 

This work focuses on a single member of a potential class of nanolaminated intermetallic compounds (IMCs), herein named the ``zigzag intermetallic'' phases or simply the ZIA phases. The first candidate ZIA phase introduced in this work is the Nb$_{3}$SiNi$_{2}$ ternary IMC, which is considered the H-phase of the Nb–-Si–-Ni system \cite{gladyshevskii1964crystal,vinicius2015thermodynamic,vinicius2019experimental,vinicius2020experimental}. This work reports the experimental synthesis of quasi-phase-pure Nb$_{3}$SiNi$_{2}$ via reactive hot pressing and also addresses, for the first time, the aspects of crystal structure nanolamination and `zigzag' atomic arrangement in the Nb$_{3}$SiNi$_{2}$ IMC. At first glance, the ZIA phases share many similarities with the MAX phases, such as crystal structure nanolamination and obeying the M$_{n+1}$AX$_n$ stoichiometric rule with n = 2; however, their true potential cannot yet be reliably assessed due to the scarcity of available data. We believe that the nanolaminated ZIA phases and their possible 2D derivatives might have a potential commensurate with that of the nanolaminated MAX phases and their 2D derivatives known as MXenes. In case future research validates this hypothesis, then a whole new research field will have been inaugurated and the prediction of Gleiter \cite{cahn1990nanostructured,gleiter1992nanostructured} on the existence of ‘nanostructured IMCs’ will have come true, at the same time. 

\section{Results and discussion}
\label{sec:results}

\subsection{Synthesis of phase-pure Nb$_{3}$SiNi$_{2}$}
\label{sec:results:item1}

\begin{figure}[hb!]
	\centering	
	\includegraphics[width=\textwidth]{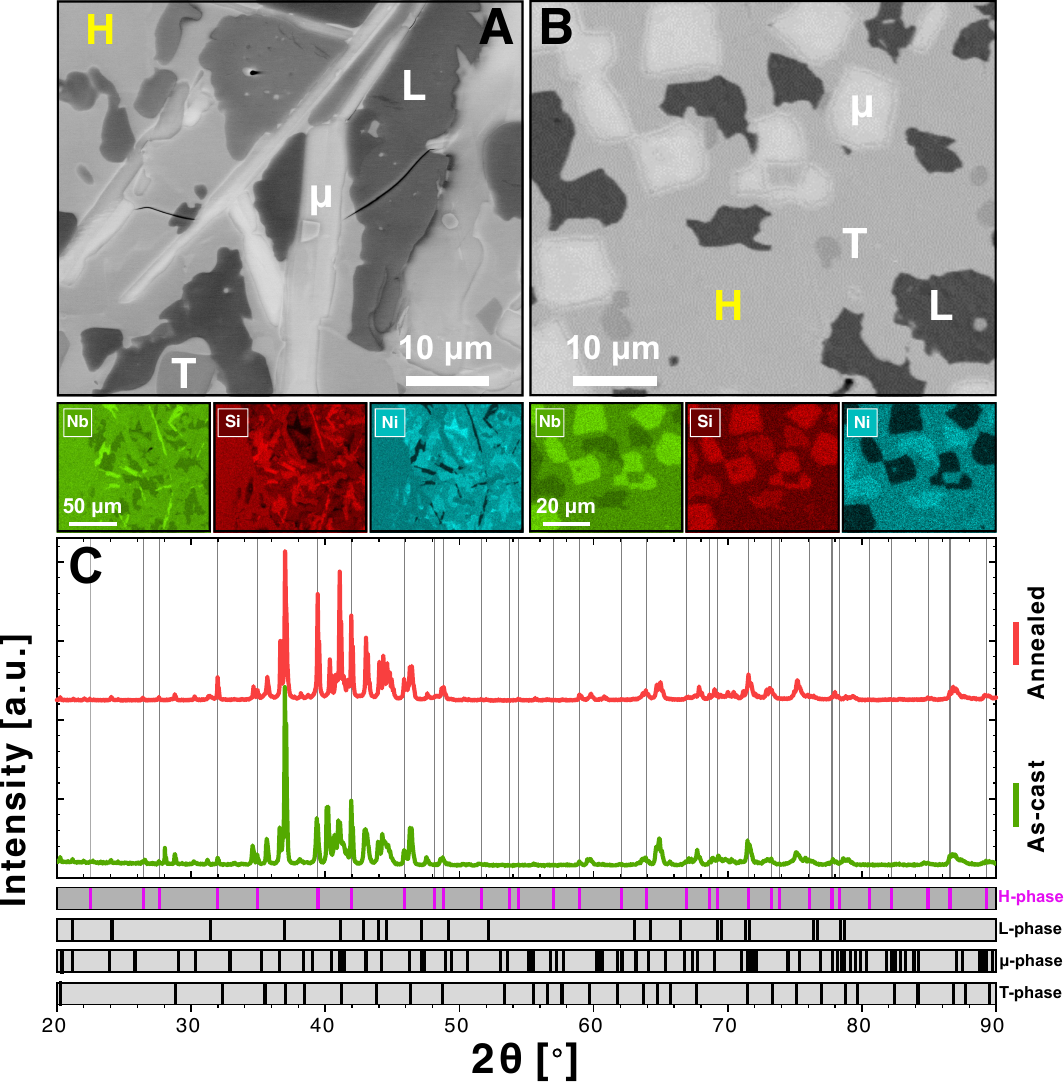}
	\caption{SEM/EDX and XRD characterization of the arc-melted Nb--Si--Ni intermetallic alloy in the as-cast state and after homogenization annealing at 1421 K for 336 h. BSE imaging and EDX elemental mapping of both as-cast (A) and annealed (B) samples show that the (thermodynamically metastable) arc-melted alloy comprises four distinct phases, \textit{i.e.}, H-Nb$_{3}$SiNi$_{2}$, L-Ni$_{3}$SiNb$_{2}$, $\mu$-Nb$_7$Ni$_6$ and T-Nb$_{4}$SiNi. (C) XRD patterns of the as-cast and annealed alloy samples confirm the presence of these four phases. The phase-of-interest, \textit{i.e.}, the ternary H-phase Nb$_{3}$SiNi$_{2}$ IMC, is highlighted in both BSE micrographs and XRD spectra.}
	\label{fig:01}
\end{figure}

\begin{figure}
	\centering	
	\includegraphics[scale=0.8]{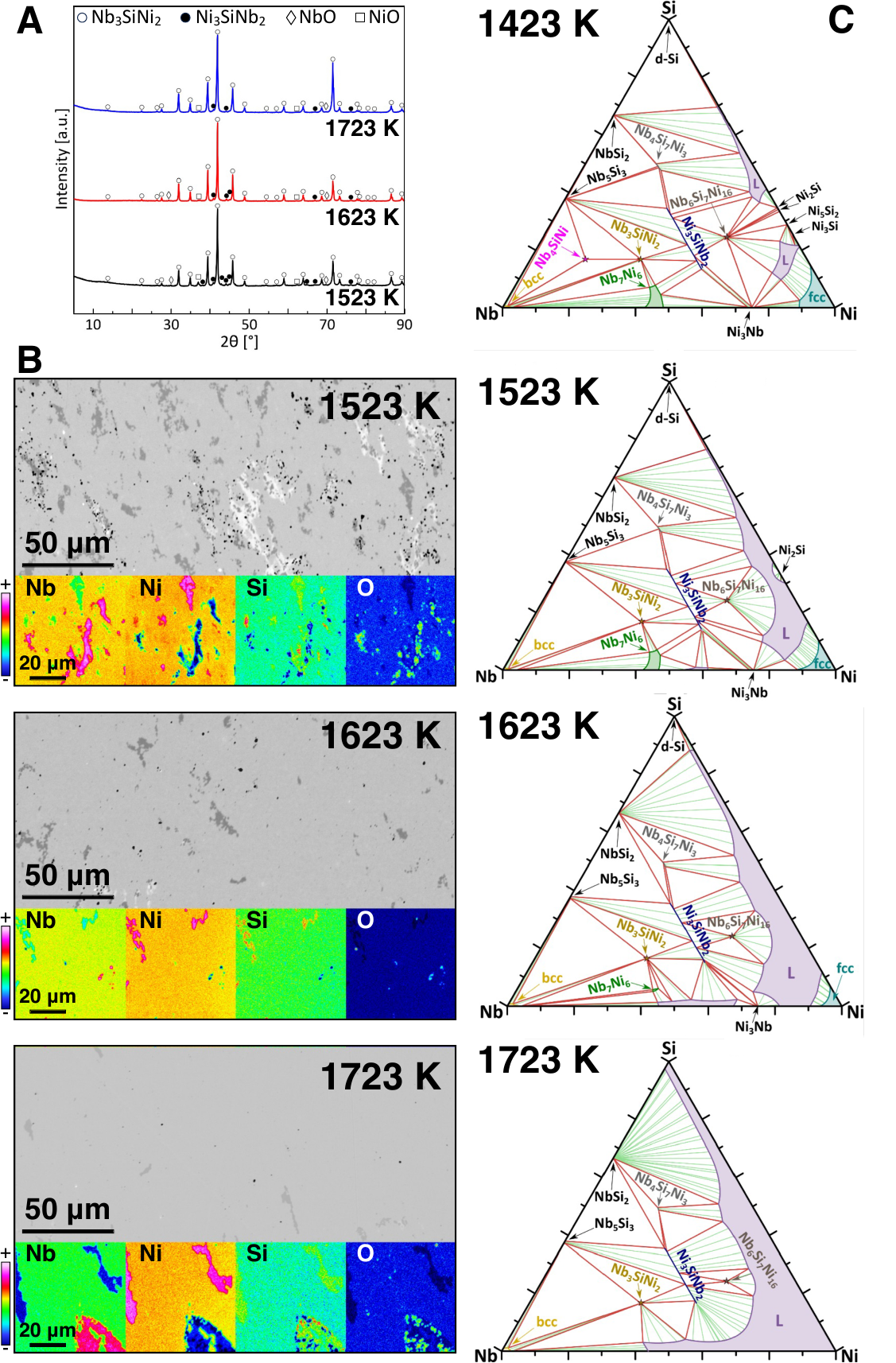}
	\caption{XRD (A) and SEM/WDXS (B) characterization of the RHP Nb--Si--Ni intermetallic alloy, showing that all alloy samples comprise mainly H-Nb$_{3}$SiNi$_{2}$ and L-Ni$_{3}$SiNb$_{2}$, with small fractions of oxidic inclusions (NbO, NiO) presumably stemming from the limited powder particle oxidation. Microstructural homogeneity and phase purity (in terms of the desired Nb$_{3}$SiNi$_{2}$ ZIA phase) seem to increase with sintering temperature. (C) CALPHAD calculations of phase equilibria in the Nb--Si--Ni system on the 1423 K, 1523 K, 1623 K and 1723 K isotherms.}
	\label{fig:02}
\end{figure}

\noindent Initially, arc melting was employed to produce bulk materials made of the Nb$_{3}$SiNi$_{2}$ IMC ZIA phase. The microstructure of the as-cast Nb–Si–Ni intermetallic alloy was first studied by SEM with the aid of a backscattered electron (BSE) detector. The compositional contrast of the BSE image in Fig. \ref{fig:01}A indicates the presence of four distinct phases in the as-cast alloy microstructure, a finding supported by both EDX elemental mapping and XRD. Phase identification by XRD (Fig. \ref{fig:01}C) revealed the (metastable) co-existence of the Nb$_{3}$SiNi$_{2}$ H-phase (candidate ZIA phase), the Nb$_{7}$Ni$_{6}$ $\mu$-phase, the Ni$_{3}$SiNb$_{2}$ ternary Laves (L) phase and the Nb$_{4}$NiSi T-phase. These four phases were peak-indexed in accordance with data provided by the ICSD database \cite{hellenbrandt2004inorganic} and were also identified as possible phases during the experimental investigation of phase equilibria in the Nb–Si–Ni refractory alloy system \cite{vinicius2020experimental}. The nominal stoichiometries and measured EDX elemental compositions of the four phases in the as-cast alloy are provided in Table \ref{res:table1}, showing that all phases formed as imperfect (off-stoichiometric) solid solutions during melt solidification rather than as stoichiometric compounds. The formation of solid solutions in the Nb--Si--Ni system has been previously reported by dos Santos \textit{et al.} \cite{vinicius2020experimental}; a strong indication of solid solution formation is encountered in the dissolution of significant amounts of Si in the Nb$_{7}$Ni$_{6}$ $\mu$-phase, as previously reported \cite{vinicius2020experimental} and also seen in Table \ref{res:table1}. Even though arc-melting targeted the formation of bulk materials (discs) made primarily of the Nb$_{3}$SiNi$_{2}$ IMC ZIA phase, the presence of four distinct phases in the as-cast alloy microstructure is a clear violation of the Gibbs' phase rule for a ternary alloy \cite{porter2009phase}, indicating that thermodynamic equilibrium has not been reached via arc melting synthesis. This is most likely caused by the loss of substantial amounts of Ni, which is volatile in vacuum above its melting point (1728 K). By looking at the CALPHAD phase equilibria calculated in this work (Fig. \ref{fig:02}C) in conjunction with the microstructure of the as-cast alloy (Fig. \ref{fig:01}A), it is reasonable to assume that the (rapid) solidification of the melt produced solid solutions of three main phases, \textit{i.e.}, the Nb$_{3}$SiNi$_{2}$ H-phase, the Nb$_{7}$Ni$_{6}$ $\mu$-phase and the Ni$_{3}$SiNb$_{2}$ L-phase, while the Ni-impoverished residual melt formed the Nb$_{4}$NiSi T-phase (see the 1423 K isotherm in Fig. \ref{fig:02}C). This hypothesis is supported by the small fraction of the Nb$_{4}$NiSi T-phase in the as-cast alloy (Fig. \ref{fig:01}A).

\begin{table}[]
\centering
\caption{Nominal stoichiometries vs. actual compositions of the phases identified by SEM/EDS in the arc-melted (as-cast \& annealed) and RHP Nb–Si–Ni intermetallic alloy samples.}
\label{res:table1}
\begin{tabular}{|llllllllll|}
\hline 
\multicolumn{10}{|c|}{\textbf{Arc-Melted Nb–Si–Ni Alloy Samples}}                                                                                                                                                                                                                             \\ \hline
\multicolumn{2}{|l|}{\textbf{Phases Identified}}   & \multicolumn{2}{l|}{\textbf{Stoichiometry}}                                                          & \multicolumn{3}{l|}{\textbf{Nb [at.\%]}}                & \multicolumn{2}{l|}{\textbf{Si [at.\%]}}          & \textbf{Ni [at.\%]} \\ \hline
\multicolumn{2}{|l|}{H-phase (as-cast)}            & \multicolumn{2}{l|}{\multirow{3}{*}{\begin{tabular}[c]{@{}l@{}}Nb$_{3}$SiNi$_{2}$\\ (ZIA phase)\end{tabular}}} & \multicolumn{3}{l|}{46.5$\pm$3.2}                           & \multicolumn{2}{l|}{18.4$\pm$4.0}                     & 35.1$\pm$1.9            \\ \cline{1-2} \cline{5-10} 
\multicolumn{2}{|l|}{H-phase (annealed)}           & \multicolumn{2}{l|}{}                                                                                & \multicolumn{3}{l|}{46.4$\pm$3.0}                           & \multicolumn{2}{l|}{18.6$\pm$3.7}                     & 35.0$\pm$1.5            \\ \cline{1-2} \cline{5-10} 
\multicolumn{2}{|l|}{H-phase (expected)}           & \multicolumn{2}{l|}{}                                                                                & \multicolumn{3}{l|}{50.0}                               & \multicolumn{2}{l|}{17.0}                         & 33.0                \\ \hline
\multicolumn{2}{|l|}{Laves phase (as-cast)}        & \multicolumn{2}{l|}{\multirow{3}{*}{Ni$_{3}$SiNb$_{2}$}}                                                       & \multicolumn{3}{l|}{31.9$\pm$2.7}                           & \multicolumn{2}{l|}{21.8$\pm$3.5}                     & 46.3$\pm$1.8            \\ \cline{1-2} \cline{5-10} 
\multicolumn{2}{|l|}{Laves phase (annealed)}       & \multicolumn{2}{l|}{}                                                                                & \multicolumn{3}{l|}{31.3$\pm$2.4}                           & \multicolumn{2}{l|}{20.1$\pm$3.1}                     & 48.6$\pm$1.2            \\ \cline{1-2} \cline{5-10} 
\multicolumn{2}{|l|}{Laves phase (expected)}       & \multicolumn{2}{l|}{}                                                                                & \multicolumn{3}{l|}{33.0}                               & \multicolumn{2}{l|}{17.0}                         & 50.0                \\ \hline
\multicolumn{2}{|l|}{$\mu$-phase (as-cast)}            & \multicolumn{2}{l|}{\multirow{3}{*}{Nb$_7$Ni$_6$*}}                                                        & \multicolumn{3}{l|}{45.2$\pm$3.1}                           & \multicolumn{2}{l|}{13.5$\pm$3.8}                     & 41.3$\pm$1.9            \\ \cline{1-2} \cline{5-10} 
\multicolumn{2}{|l|}{$\mu$-phase (annealed)}           & \multicolumn{2}{l|}{}                                                                                & \multicolumn{3}{l|}{43.8$\pm$3.0}                           & \multicolumn{2}{l|}{13.5$\pm$3.7}                     & 42.7$\pm$1.5            \\ \cline{1-2} \cline{5-10} 
\multicolumn{2}{|l|}{$\mu$-phase (expected)}           & \multicolumn{2}{l|}{}                                                                                & \multicolumn{3}{l|}{53.8}                               & \multicolumn{2}{l|}{-}                            & 46.2                \\ \hline
\multicolumn{2}{|l|}{T-phase (as-cast)}            & \multicolumn{2}{l|}{\multirow{3}{*}{Nb$_4$NiSi}}                                                        & \multicolumn{3}{l|}{62.8$\pm$3.8}                           & \multicolumn{2}{l|}{25.8$\pm$4.7}                     & 11.4$\pm$2.5            \\ \cline{1-2} \cline{5-10} 
\multicolumn{2}{|l|}{T-phase (annealed)}           & \multicolumn{2}{l|}{}                                                                                & \multicolumn{3}{l|}{62.6$\pm$3.8}                           & \multicolumn{2}{l|}{25.3$\pm$4.7}                     & 12.0$\pm$1.9            \\ \cline{1-2} \cline{5-10} 
\multicolumn{2}{|l|}{T-phase (expected)}           & \multicolumn{2}{l|}{}                                                                                & \multicolumn{3}{l|}{66.0}                               & \multicolumn{2}{l|}{17.0}                         & 17.0                \\ \hline
\multicolumn{10}{|c|}{\textbf{RHP Nb–Si–Ni Alloy Samples}}                                                                                                                                                                                                                                    \\ \hline
\multicolumn{1}{|l|}{\textbf{Sintering T [K]}} & \multicolumn{2}{l|}{\textbf{Phases Identified}}      & \multicolumn{2}{l|}{\textbf{Nb [at.\%]}}            & \multicolumn{1}{l|}{\textbf{Si [at.\%]}} & \multicolumn{2}{l|}{\textbf{Ni [at.\%]}} & \multicolumn{2}{l|}{\textbf{O [at.\%]}}  \\ \hline
\multicolumn{1}{|l|}{\multirow{3}{*}{1523}}     & \multicolumn{2}{l|}{H-Nb$_{3}$SiNi$_{2}$}                      & \multicolumn{2}{l|}{53.86$\pm$0.16}                     & \multicolumn{1}{l|}{12.71$\pm$0.03}          & \multicolumn{2}{l|}{33.43$\pm$0.12}          & \multicolumn{2}{l|}{-}                   \\ \cline{2-10} 
\multicolumn{1}{|l|}{}                          & \multicolumn{2}{l|}{L-Ni$_{3}$SiNb$_{2}$}                      & \multicolumn{2}{l|}{38.60$\pm$0.12}                     & \multicolumn{1}{l|}{17.63$\pm$0.03}          & \multicolumn{2}{l|}{43.77$\pm$0.12}          & \multicolumn{2}{l|}{-}                   \\ \cline{2-10} 
\multicolumn{1}{|l|}{}                          & \multicolumn{2}{l|}{$\mu$-Nb$_7$Ni$_6$}                        & \multicolumn{2}{l|}{71.99$\pm$0.13}                     & \multicolumn{1}{l|}{17.17$\pm$0.03}          & \multicolumn{2}{l|}{10.85$\pm$0.07}          & \multicolumn{2}{l|}{-}                   \\ \hline
\multicolumn{1}{|l|}{\multirow{3}{*}{1623}}     & \multicolumn{2}{l|}{H-Nb$_{3}$SiNi$_{2}$}                      & \multicolumn{2}{l|}{53.57$\pm$0.15}                     & \multicolumn{1}{l|}{13.26$\pm$0.03}          & \multicolumn{2}{l|}{33.17$\pm$0.12}          & \multicolumn{2}{l|}{-}                   \\ \cline{2-10} 
\multicolumn{1}{|l|}{}                          & \multicolumn{2}{l|}{L-Ni$_{3}$SiNb$_{2}$}                      & \multicolumn{2}{l|}{36.25$\pm$0.15}                     & \multicolumn{1}{l|}{17.49$\pm$0.04}          & \multicolumn{2}{l|}{46.27$\pm$0.16}          & \multicolumn{2}{l|}{-}                   \\ \cline{2-10} 
\multicolumn{1}{|l|}{}                          & \multicolumn{2}{l|}{$\mu$-Nb$_7$Ni$_6$}                        & \multicolumn{2}{l|}{-}                              & \multicolumn{1}{l|}{-}                   & \multicolumn{2}{l|}{-}                   & \multicolumn{2}{l|}{-}                   \\ \hline
\multicolumn{1}{|l|}{\multirow{3}{*}{1723}}     & \multicolumn{2}{l|}{H-Nb$_{3}$SiNi$_{2}$}                      & \multicolumn{2}{l|}{52.80$\pm$0.17}                     & \multicolumn{1}{l|}{13.12$\pm$0.04}          & \multicolumn{2}{l|}{34.08$\pm$0.14}          & \multicolumn{2}{l|}{-}                   \\ \cline{2-10} 
\multicolumn{1}{|l|}{}                          & \multicolumn{2}{l|}{L-Ni$_{3}$SiNb$_{2}$}                      & \multicolumn{2}{l|}{36.32$\pm$0.11}                     & \multicolumn{1}{l|}{17.47$\pm$0.03}          & \multicolumn{2}{l|}{46.21$\pm$0.11}          & \multicolumn{2}{l|}{-}                   \\ \cline{2-10} 
\multicolumn{1}{|l|}{}                          & \multicolumn{2}{l|}{$\mu$-Nb$_7$Ni$_6$}                        & \multicolumn{2}{l|}{65.96$\pm$0.21}                     & \multicolumn{1}{l|}{8.43$\pm$0.04}           & \multicolumn{2}{l|}{4.65$\pm$0.09}           & \multicolumn{2}{l|}{20.96$\pm$0.05}          \\ \hline
\end{tabular}
\\ \textbf{*} Si has been reported to dissolve in the binary Nb$_7$Ni$_6$ IMC, forming $\mu$-phase solid solutions \cite{vinicius2020experimental}.
\end{table}

The homogenization annealing at 1421 K was performed, first and foremost, to investigate whether the four-phase assembly of the as-cast Nb–Si–Ni intermetallic alloy would converge to an alloy with a higher fraction of the Nb$_{3}$SiNi$_{2}$ IMC ZIA phase. The annealing aimed also at a first assessment of the thermal stability of the Nb$_{3}$SiNi$_{2}$ ZIA phase of interest in this work. The SEM/EDS/XRD analysis of the annealed alloy (Figs. \ref{fig:01}B \& 1C) showed that it still consisted of the same four phases with a slightly increased fraction of the Nb$_{3}$SiNi$_{2}$ H-phase that appears able to sustain these conditions; the increase of the latter was not sufficient to suggest that the selected annealing conditions were capable of transforming the phase assembly of the as-cast alloy into an alloy primarily composed of the Nb$_{3}$SiNi$_{2}$ IMC. This suggests that solid-state diffusion at 1421 K is relative slow, and that annealing treatments much longer than 336 h would be required to achieve alloy homogenization at 1421 K; moreover, the fact that the alloy composition after solidification of the melt appears to be off the Nb$_{3}$SiNi$_{2}$ stoichiometry (primarily due to the loss of Ni) will prevent the formation of a phase-pure Nb$_{3}$SiNi$_{2}$ material, irrespective of the annealing duration and temperature. Fig. \ref{fig:01}B reveals that the microstructure of the annealed alloy was slightly modified, suggesting limited coarsening and change of grain morphology (\textit{e.g.}, the Nb$_7$Ni$_6$ $\mu$-phase grains appear more angular and equiaxed in the annealed alloy than in the as-cast alloy). The limited microstructural changes indicate that the thermodynamically metastable as-cast alloy strives to achieve equilibrium during isothermal annealing, forming more Nb$_{3}$SiNi$_{2}$ and transitioning towards a slightly coarser microstructure, however, the imposed annealing conditions are not conducive to achieving equilibrium. The elemental compositions of the four phases in the annealed alloy, as determined by EDX analysis, are provided in Table \ref{res:table1}. Additional BSE images of the microstructures of both as-cast and annealed arc-melted samples are given in Fig. S1 (Supplementary Information). The unsuccessful attempt to synthesize phase-pure Nb$_{3}$SiNi$_{2}$ via arc melting showed that this processing route is difficult to master, mainly due to the poorly controlled losses of the constituting elements above their melting points (\textit{e.g.}, T$_m$(Ni) = 1728 K, T$_m$(Nb) = 2750 K, T$_m$(Si) = 1683 K). Moreover, it became clear that it is necessary to calculate the equilibrium phase mixtures at the annealing temperature (1421 K) and at higher temperatures (in order to understand melt solidification), as prior studies have only examined the 1073, 1323 and 1473 K isotherms of the Nb--Si--Ni system \cite{gladyshevskii1964crystal,vinicius2015thermodynamic,vinicius2019experimental,vinicius2020experimental}.

The results of this work contradict the early work of Gladyshevskii \textit{et al.} \cite{gladyshevskii1964crystal}, who reported that by melting a feedstock of elemental powders under an inert gas atmosphere  (presumably, argon), they were able to produce quasi-phase-pure Nb$_{3}$SiNi$_{2}$ with only a small amount of a ternary Laves phase (presumably, the Ni$_{3}$SiNb$_{2}$ ternary IMC also detected in this work). These authors used light optical microscopy to confirm the phase purity of their arc-melted alloys, and powder XRD to propose the existence of a group of (atomically ordered) ternary IMCs with the H-phase structure (space group $Fd\bar{3}m$) and the R$_{3}$SiNi$_{2}$ general stoichiometry, with R being the early transition metals Mn, Cr, V, Nb and Ta \cite{gladyshevskii1964crystal}. Apart from demonstrating a much better command of the arc-melting processing route, the 1964 pioneering work by Gladyshevskii \textit{et al.} \cite{gladyshevskii1964crystal} paved the way for the study of the ZIA phases proposed in this work. In a similar manner, the 1967 work by Jeitschko \textit{et al.} \cite{jeitschko1963kohlenstoffhaltige} opened the way for the further development of the class of early transition metal ternary carbides/nitrides known as the MAX phases almost 40 years later by Barsoum \textit{et al.} \cite{barsoum1996synthesis,barsoum2001max}. 

A thorough thermodynamic assessment of the Nb--Si--Ni system was only recently conducted – 50 years after Gladyshevskii’s pioneering work – in a set of works published by dos Santos \textit{et al.} \cite{vinicius2015thermodynamic,vinicius2019experimental,vinicius2020experimental}. In an attempt to reproduce the results of Gladyshevskii \textit{et al.} \cite{gladyshevskii1964crystal}, dos Santos \textit{et al.} \cite{vinicius2019experimental} synthesized 12 different alloy compositions in the Nb--Si--Ni system via arc melting and subsequently homogenized the as-cast samples at 1073 K for longer times (1000 h). Curiously, the Nb$_{3}$SiNi$_{2}$ phase was not found in any of the 12 homogenized samples characterized by means of SEM/EDX and XRD. It is presently not understood whether the difference in atmosphere during melting – reducing (H$_{2}$) in Gladyshevskii \textit{et al.} \cite{gladyshevskii1964crystal} and possibly slightly oxidizing (argon, Ar, of unknown purity) in dos Santos \textit{et al.} \cite{vinicius2015thermodynamic,vinicius2019experimental,vinicius2020experimental} – is accountable for the difference in the produced arc-melted alloys. Based on their data, dos Santos \textit{et al.} suggested a homogenization isothermal annealing at higher temperatures to explore the thermodynamic equilibrium in this system \cite{vinicius2019experimental}. In a more extensive follow-up study, dos Santos \textit{et al.} investigated 48 alloy compositions in the Nb--Si--Ni system, implementing two different heat treatment approaches: (a) 1323 K for 336 h and (b) 1473 K for 504 h \cite{vinicius2020experimental}. Using SEM/EDX and XRD after homogenization annealing, the Nb$_{3}$SiNi$_{2}$ phase was only observed in seven samples, but none of these were phase-pure Nb$_{3}$SiNi$_{2}$. All 7 samples exhibited triphasic microstructures, each comprising two or three of the four principal phases, \textit{i.e.}, the ternary Laves, $\mu$, T- and H-phases; small fractions of the NbNi$_{3}$ IMC or the unknown Z phase were also reported in 3 of the 7 samples. The Z phase is one of three phases with unresolved crystal structure – tentatively designated X, Y and Z – found by dos Santos \textit{et al.} after annealing select Nb-Si-Ni alloy compositions at 1473 K for 504 h \cite{vinicius2020experimental}. 

Careful consideration of both the prior attempts by dos Santos \textit{et al.} \cite{vinicius2019experimental,vinicius2020experimental} and the outcome of this work demonstrated that the synthesis of phase-pure Nb$_{3}$SiNi$_{2}$ via arc melting is challenging. Even though Gladyshevskii \textit{et al.} \cite{gladyshevskii1964crystal} reported that the production of quasi-phase-pure Nb$_{3}$SiNi$_{2}$ is possible via the melting of an elemental powder feedstock with identical composition in an inert gas atmosphere, high-vacuum plasma arc melting methods have been empirically known to face challenges during the synthesis of compounds with strict stoichiometries, such as the ternary IMCs designated as the ZIA phases in this work, or the ternary carbides/nitrides known as the MAX phases \cite{coulombe2000theoretical}. As already mentioned, one important challenge is the evaporation of alloying elements in the vacuum chamber \cite{radhakrishnan1999synthesis}. Given the strict stoichiometry of the ZIA phases and the 312 MAX phases, the processing route as well as the starting powder feedstock must be always carefully adjusted to produce materials with high phase purity. For example, Goossens \textit{et al.} \cite{goossens2022synthesis} explained rather recently why the use of early transition metal hydride powders produces MAX phase-based ceramics with higher phase purities than the use of elemental starting powders. These authors proved that the use of metal hydride powders lowers the formation temperature of MAX phase precursor phases (\textit{e.g.}, binary IMCs), increases the formation temperature of competing phases (\textit{e.g.}, binary carbides), and suppresses the oxidation of oxidation-prone powders (\textit{e.g.}, Zr) by providing a reducing (H$_{2}$-rich) atmosphere caused by powder dehydrogenation, thus allowing the use of finer powder feedstocks. Based on the above, it was decided in this work to employ reactive hot pressing (RHP) of carefully adjusted powder feedstocks consisting of hydride (NbH$_{0.89}$) and elemental powders (Ni \& Si) to produce phase-pure Nb$_{3}$SiNi$_{2}$. The synthesis efforts proved to be successful at three different sintering temperatures (\textit{i.e.}, 1523 K, 1623 K, and 1723 K), producing quasi-phase-pure Nb$_{3}$SiNi$_{2}$ with a small fraction of the ternary Ni$_{3}$SiNb$_{2}$ Laves phase, and tiny amounts of oxidic inclusions (NbO \& NiO), presumably stemming from the limited oxidation of the raw powders. BSE images of the microstructures and WDXS elemental maps of all quasi-phase-pure Nb$_{3}$SiNi$_{2}$ materials are shown in Fig. \ref{fig:02}B, next to their XRD patterns (Fig. \ref{fig:02}A). Additional BSE micrographs and EDS elemental maps of all RHP alloy samples are shown in Fig. S2 (Supplementary Information). 

The results of CALPHAD calculations of the phase equilibria in the ternary Nb--Si--Ni system at 1423 K (\textit{i.e.}, close to the temperature of isothermal annealing, 1421 K, of the arc-melted alloy), 1523 K, 1623 K and 1723 K are shown in Fig. \ref{fig:02}C. These calculations suggest that melt solidification during vacuum arc melting produced mainly a ternary phase mixture comprising the L-Ni$_{3}$SiNb$_{2}$ line compound, the H-Nb$_{3}$SiNi$_{2}$ point compound (targeted ZIA phase), and the $\mu$-Nb$_7$Ni$_6$ phase-field compound with enhanced Si solid solubility. Due to the anticipated loss of Ni during vacuum arc melting, a small fraction of the melt was Ni-poor and solidified as the T-Nb$_{4}$SiNi point compound, which appears only in the 1423 K isotherm (Fig. \ref{fig:02}C). On the other hand, all three alloys synthesized via RHP at 1523/1623/1723 K resulted in quasi-binary phase mixtures made of the H-Nb$_{3}$SiNi$_{2}$ (targeted ZIA phase) and L-Ni$_{3}$SiNb$_{2}$ compounds. The NbO and NiO oxide impurities present in the RHP alloys are not encountered in the arc-melted alloys due to the much higher processing temperatures associated with the complete melting of all raw elemental powders during arc melting synthesis. 

\newpage

\subsection{TEM/STEM characterization of the Nb$_{3}$SiNi$_{2}$}
\label{sec:results:item2}

\begin{figure}[hb!]
	\centering	
	\includegraphics[width=\textwidth]{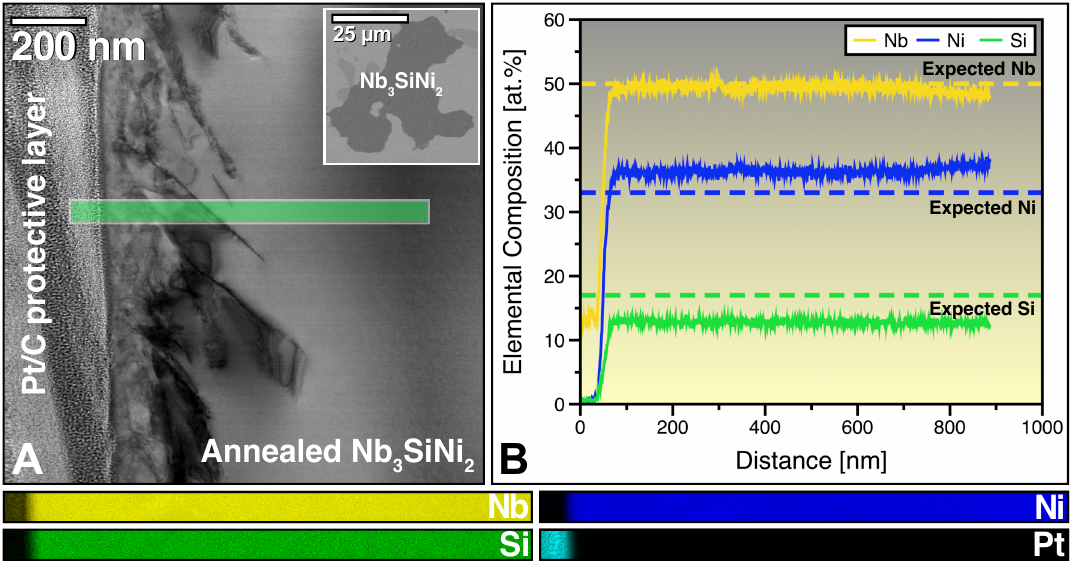}
	\caption{STEM/EDX analysis of a FIB foil lifted-out from a large Nb$_{3}$SiNi$_{2}$ IMC grain in the annealed arc-melted alloy. (A) BF-STEM micrograph of the Nb$_{3}$SiNi$_{2}$ IMC grain showing sub-surface mechanical damage caused by sample grinding/polishing. The inset is a BSE image of the Nb$_{3}$SiNi$_{2}$ grain selected for this FIB foil lift-out. (B) STEM/EDS analysis of the area inside the green rectangle of Fig. 3A shows the homogeneous distribution of Nb, Si and Ni (Nb/Si/Ni elemental maps, bottom), confirming that the average chemical composition of this grain within the analyzed area deviates slightly from the Nb$_{3}$SiNi$_{2}$ stoichiometry (Nb/Si/Ni profiles, top).}
	\label{fig:03}
\end{figure}

\noindent Even though arc melting produced a bulk sample with a low Nb$_{3}$SiNi$_{2}$ content, FIB was used to lift out thin foils from sufficiently large Nb$_{3}$SiNi$_{2}$ grains in the annealed sample (Fig. \ref{fig:03}A) for characterization of the Nb$_{3}$SiNi$_{2}$ ZIA phase by means of (S)TEM/SAED. This characterization was decided before quasi-phase-pure bulk samples were produced via RHP, so as to gain a first in-depth understanding of the crystal structure of the Nb$_{3}$SiNi$_{2}$ ZIA phase. A bright-field STEM (BF-STEM) cross-sectional micrograph of such a Nb$_{3}$SiNi$_{2}$ grain is shown in Fig. \ref{fig:03}A. The defects observed within the first 500 nm beneath the sample surface (protected by a combined Pt/C strip during FIB lift-out) are attributed to mechanical damage induced during grinding/polishing of the bulk annealed sample. The presence of sub-surface mechanical damage did not affect the chemistry of the analyzed grain, which was found by STEM/EDS to be compositionally homogeneous and close to the Nb$_{3}$SiNi$_{2}$ stoichiometry with limited deviations from the expected Ni (5 at.\% excess) and Si (5 at.\% deficiency) contents (Fig. \ref{fig:03}B).

\begin{figure}
	\centering	
	\includegraphics[width=\textwidth]{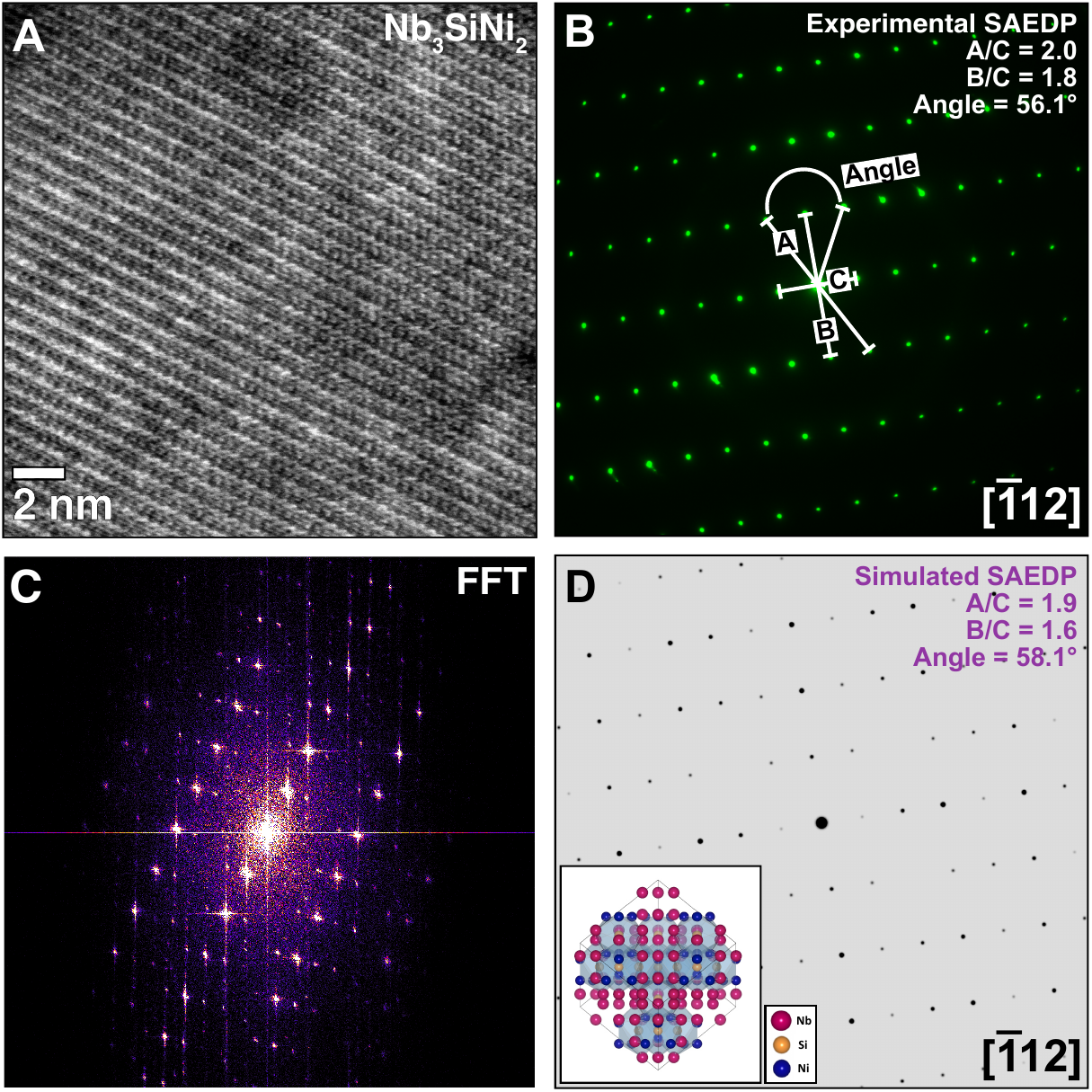}
	\caption{STEM/SAED characterization of the single Nb$_{3}$SiNi$_{2}$ IMC grain of Figure 3A. (A) High-resolution BF-STEM micrograph of the Nb$_{3}$SiNi$_{2}$ IMC grain viewed in the low-index $[\bar{1}12]$ zone axis. (B) Experimental SAEDP and (C) Fast Fourier Transform (FFT) pattern of the $[\bar{1}12]$ zone axis of the Nb$_{3}$SiNi$_{2}$ IMC. (D) Simulated SAEDP of the $[\bar{1}12]$ zone axis of the Nb$_{3}$SiNi$_{2}$ IMC; the inset presents the unit cell of the Nb$_{3}$SiNi$_{2}$ in the $[\bar{1}12]$ zone axis. The simulated SAEDP was produced using SingleCrystal \cite{palmer2015visualization} and experimental data that are available in the ICSD database \cite{gladyshevskii1964crystal}. The experimental and simulated SAEDPs of Figures 4B and 4D, respectively, confirm that the Nb$_{3}$SiNi$_{2}$ candidate ZIA phase has the diamond cubic crystal structure (\textit{fcc} Bravais lattice; $Fd\bar{3}m$; space group 227).}
	\label{fig:04}
\end{figure}

Given the rather satisfactory chemistry match with the targeted Nb$_{3}$SiNi$_{2}$ stoichiometry, further characterization of the same Nb$_{3}$SiNi$_{2}$ grain by STEM \& SAED (selected area diffraction) was deemed necessary to elucidate the crystal structure. Figure \ref{fig:04}A  is an atomically resolved BF-STEM image of the Nb$_{3}$SiNi$_{2}$ ZIA phase, showing the nanolayered atomic arrangement in a ``zigzag'' pattern, which was observed in specific low-index crystallographic orientations. Figure \ref{fig:04}B shows a SAEDP recorded after orienting the sample onto the low-index [$\bar{1}12$] zone axis. Using published crystallographic data (CIF file code ICSD-2044323 \cite{gladyshevskii1964crystal}) and the SingleCrystal software package \cite{palmer2015visualization}, a kinematic simulation of the SAEDP (Fig. \ref{fig:04}D) was made and compared with the recorded SAEDP (Fig. \ref{fig:04}B). Experimental and simulated SAEDPs were in excellent agreement, whilst all collected data suggested that the crystal structure of the experimentally synthesized Nb$_{3}$SiNi$_{2}$ is diamond cubic ($Fd\bar{3}m$, space group 227) \cite{gladyshevskii1964crystal}. This crystal structure characterizes not only the ternary H-phase (Nb$_{3}$SiNi$_{2}$) of the Nb--Si--Ni system \cite{gladyshevskii1964crystal,vinicius2019experimental,vinicius2020experimental}, but also the ternary IMCs with identical stoichiometry in the Mn--Si--Ni (Mn$_{3}$SiNi$_{2}$) \cite{gladyshevskii1964crystal,kuo1986friauf,kolenda1990magnetic}, V--Si--Ni (V$_{3}$SiNi$_{2}$) \cite{gladyshevskii1964crystal}, Cr--Si--Ni (Cr$_{3}$SiNi$_{2}$) \cite{gladyshevskii1964crystal}, Ta--Si--Ni (Ta$_{3}$SiNi$_{2}$) \cite{gladyshevskii1964crystal,wang2021experimental}, Nb--Si--Co (Nb$_{3}$SiCo$_{2}$) \cite{kuz1964crystal}, Na--Au--In (Na$_{3}$AuIn$_{2}$) \cite{li2005participation}, Na--Ag--In (Na$_{3}$AgIn$_{2}$) \cite{li2005participation}, and Mg--Ga--Ni (Mg$_{3}$GaNi$_{2}$) \cite{pavlyuk2020new} systems, apparently being the preferred crystal structure for the entire class of the (312) ZIA phases, just like the hexagonal close-packed (\textit{hcp}; $P6_{3}/mmc$, space group 194) crystal structure encompasses all MAX phase compounds \cite{barsoum2001max,goossens2021microstructure}. Having said that, it is also worthwhile adding that few IMCs with the 312 stoichiometry and the \textit{hcp} ($P6_{3}/mmc$, space group 194) crystal structure have been sporadically identified, possibly suggesting the existence of a not yet fully identified and/or explored overarching trend in the formation of nanolaminated solids, such as the known MAX phases and the herein proposed ZIA phases. Examples of ternary 312 IMCs with the \textit{hcp} crystal structure have been previously reported in the Co--Si--Nb (Co$_3$SiNb$_2$) \cite{kuz1964crystal}, Ni--Si--Nb (Nb$_{3}$SiNi$_{2}$) \cite{gladyshevskii1964crystal} and Cu--Si--Mg (Cu$_3$SiMg$_2$) \cite{kuz1964crystal} systems. In view of the scarcity of available data, it would be premature to speculate on the relation between the \textit{fcc} ternary IMCs identified in this work as the ZIA phases and the largely unexplored \textit{hcp} ternary IMCs, or to assume that the latter are also characterized by crystal structure nanolamination, even though such a hypothesis seems quite plausible.

\subsection{Opportunities arising from the advent of the ZIA phases}
\label{sec:results:item3}

\noindent In the search for new refractory materials with exceptional properties, both ZIA and MAX phases are nanolaminated crystalline solids that share the same historical roots, dating back to the advent of space exploration in the 1960s \cite{jeitschko1963kohlenstoffhaltige,jeitschko1967kristallstruktur,radovic2013max,gladyshevskii1964crystal,vinicius2019experimental,vinicius2020experimental}. Despite the fact that they have been both discovered in the same period of time, there is very little, if any, knowledge of the properties and possible applications of the ternary IMC ZIA phases and their (not yet synthesized) 2D derivatives, as opposed to the extensively studied MAX phase ternary carbides/nitrides, their higher order solid solutions and 2D derivatives known as MXenes. Therefore, this section will attempt to describe the basic differences and similarities of the rather mature MAX phases vs. the emerging ZIA phases, discussing potential applications of the latter based on the experience gained through the MAX-phase research over the past 30 years.

\begin{figure}
	\centering	
	\includegraphics[width=\textwidth]{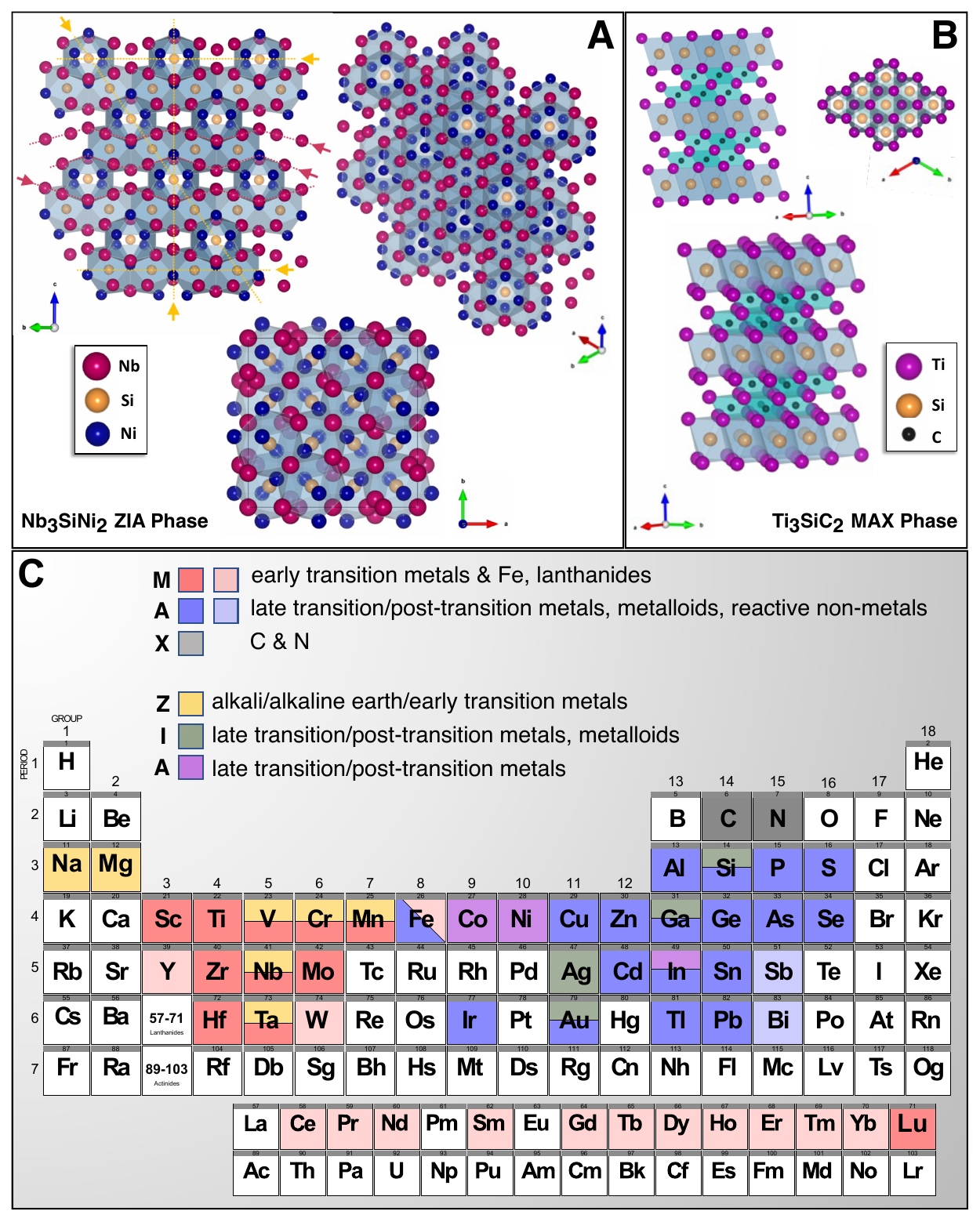}
	\caption{Comparison of the unit cells of (A) the herein synthesized and characterized \textit{fcc} Nb$_{3}$SiNi$_{2}$ ZIA phase with (B) the well-known \textit{hcp} Ti$_{3}$SiC$_{2}$ MAX phase \cite{jeitschko1967kristallstruktur,barsoum1996synthesis}. Fig. 5A  (bottom) shows 1 unit cell (96 atoms/unit cell) of the Nb$_{3}$SiNi$_{2}$ ZIA phase, as well as the [110] (upper left) and [111] (upper right) zone axes using 4 unit cells. Red arrows pinpoint atomic zigzagging in the [110] zone axis, yellow arrows indicate crystal structure nanolamination. Fig. 5B (bottom) shows 8 unit cells of the Ti$_{3}$SiC$_{2}$ MAX phase (6 atoms/unit cell), as well as the $[11\bar{2}0]$ (upper left) and [0001] (upper right) zone axes. (C) The ZIA phases extend the concept of crystal structure nanolamination beyond the early transition metal carbides/nitrides known as the MAX phases. Together, the two classes of nanolaminated solids span almost the whole period table, comprising jointly elements from groups 1-16. The M-elements of the MAX phases are shown using 2 shades of red: the darker indicates full occupation of the M-site, the lighter indicates partial occupation (solid solutions). The same holds for the 2 shades of blue dedicated to the A-elements of the MAX phases. Note: the periodic table in Fig. 5C is a modified version of the original American Chemical Society's periodic table.}
	\label{fig:05}
\end{figure}

Using data compiled by the Materials Genome Project \cite{jain2013commentary}, the first candidate ZIA phase, \textit{i.e.}, the Nb$_{3}$SiNi$_{2}$ IMC, is herein discussed with respect to the extensively studied Ti$_{3}$SiC$_{2}$ MAX phase. First, they are both ternary compounds that belong to different elemental systems, \textit{i.e.}, Nb$_{3}$SiNi$_{2}$ forms in the Nb–Si–Ni system, while Ti$_{3}$SiC$_{2}$ in the Ti--Si--C system. Even though both compounds obey the 312 MAX phase stoichiometric rule, i.e., M$_{n+1}$AX$_{n}$ with n = 2, Nb$_{3}$SiNi$_{2}$ is clearly an IMC, while Ti$_{3}$SiC$_{2}$ is classified as a ceramic. Moreover, there is a remarkable difference in crystal structure complexity, with the unit cell of the Nb$_{3}$SiNi$_{2}$ ZIA phase (Fig. \ref{fig:05}A) being significantly larger than the unit cell of the Ti$_{3}$SiC$_{2}$ MAX phase (Fig. \ref{fig:05}B). More specifically, the Ti$_{3}$SiC$_{2}$ unit cell is made of 6 atoms, while the Nb$_{3}$SiNi$_{2}$ one is made of 96 atoms, resulting in unit cell volumes that differ in size by one order of magnitude (Table S1 in the Supplementary Information). Both ternary compounds are characterized by centrosymmetric close-packed Bravais lattices, \textit{i.e.}, the Ti$_{3}$SiC$_{2}$ MAX phase is \textit{hcp} ($P6_{3}/mmc$, space group 194) whilst the Nb$_{3}$SiNi$_{2}$ ZIA phase is \textit{fcc} (face-centered cubic; $Fd\bar{3}m$, space group 227); apparently, such close-packed atomic configurations seem to favor crystal structure nanolamination.


A groundbreaking impact of the herein proposed ZIA phases is that they actually extend the concept of crystal structure nanolamination much beyond the early transition metal carbides/nitrides known as the MAX phases. As may be seen in the color-coded periodic table of elements of Fig. \ref{fig:05}C, the M-elements used to form ternary MAX phases are early transition metals (groups 3-7), the A-elements belong mainly to groups 13-15 (post-transition metals \& metalloids) with limited expansion to groups 8-12 (late transition metals) and group 16 (reactive non-metals), while the X-element is either C or N (reactive non-metals in groups 14 and 15, respectively). On the other hand, the ZIA IMCs that have been hitherto identified utilize Z-elements from groups 1 (alkali metals) and 2 (alkaline earth metals) and groups 5-7 (early transition metals), I-elements from group 11 (late transition metals) and groups 13 and 14 (post-transition metals \& metalloids), while the A-elements belong either to groups 9 and 10 (late transition metals) or group 13 (the post-transition metal indium, In). In short, the synthesis of nanolaminated crystalline solids (MAX and ZIA phases) requires different combinations of constituent elements from groups 1-16 in the periodic table. This heralds the potential fabrication of ternary compounds and higher-order solid solutions with unique properties and very broad applicability, not only in the form of the 3D ZIA and MAX phases, but also in the form of their 2D derivatives, which have so far only been explored for the MAX phases in the form of the rapidly developing MXenes. Especially the discovery of ZIA phases comprising elements that have hitherto not been introduced in the MAX phases (\textit{e.g.}, alkali and alkaline earth metals, several late transition metals) paves the way for the fabrication of new nanolaminated crystalline solids with certain properties (\textit{e.g.}, ferromagnetic properties, enhanced/tailorable ductility, superconductivity, etc.) that are typically not associated with the MAX phases. This hypothesis finds support in the work of Kolenda \textit{et al.} \cite{kolenda1990magnetic}, who reported on the magnetic properties of the Mn$_{3}$SiNi$_{2}$ candidate ZIA phase. The new range of properties achieved by the ZIA phases is also expected to be reflected in their 2D derivatives (assuming these can first be produced experimentally), probably giving rise to the next generation of 2D materials for use in flexible and portable microelectronics, batteries, sensors, etc.

It is worthwhile noting that in all hitherto reported \textit{fcc} ternary IMCs now considered as candidate ZIA phases, the electronegativity, $\chi$, of the Z-element is invariably smaller than the electronegativities of both I- and A-elements (Table S2 in the Supplementary Information). Moreover, the atomic radius of the Z-element is larger than the atomic radii of both I- and A-elements with the exception of Cr$_{3}$SiNi$_{2}$, where the atomic radius (125 pm) of the Z-element, Cr, is equal to that of the A-element, Ni (Table S2 in the Supplementary Information). Interestingly, the structurally unexplored 312 \textit{hcp} ternary IMCs Co$_3$SiNb$_2$ and Ni$_3$SiNb$_2$ could be considered as `compositionally inverted' Nb$_3$SiCo$_2$ and Nb$_3$SiNi$_2$ ZIA phases, respectively, produced by interchanging the Z- and A-elements while still respecting the strict stoichiometric rule of the 312 MAX phases. For these two \textit{hcp} IMCs as well as for the Cu$_3$SiMg$_2$ one, the electronegativity of the third element (Nb or Mg) is smaller than the electronegativities of the first two elements (Co, Ni or Cu). Also, the atomic radius of the third element (Nb or Mg) is larger than the atomic radii of the first two elements (Co, Ni or Cu). Based on these structural considerations and the fact that the \textit{hcp} ternary IMCs Co$_3$SiNb$_2$, Ni$_3$SiNb$_2$ and Cu$_3$SiMg$_2$ are characterized by the same hexagonal close-packed lattice ($P6_{3}/mmc$, space group 194) as the MAX phases, one could indeed expect crystal structure nanolamination in these ternary IMCs as well. This hypothesis, however, merits further dedicated investigation.

A unique feature of the ZIA phases is that the I- and A-elements could be chosen from the same group in the periodic table. For example, both the A-element (In) in Na$_3$AuIn$_2$, and Na$_3$AgIn$_2$ as well as the I-element (Ga) in Mg$_3$GaNi$_2$ belong to group 13. This freedom in the choice of the elements making up the ZIA phases does not characterize the MAX phases, where only two X-elements are allowed (\textit{i.e.}, C or N), and these are not known to be replaced by any of the candidate M- and A-elements. The only known exception of an element that has been rather flexible in occupying the M- or A-site of specific MAX phase compounds is the late transition metal iron (Fe), which belongs to group 8 of the periodic table. More specifically, Fe has been fairly recently reported to form the (Ti$_{1-x}$,Fe$_x$)$_3$AlC$_2$ solid solutions on the M-site \cite{sun2021magnetic}, or to occupy fully the A-site of the ternary MAX phases Ta$_2$FeC, Ti$_2$FeN and Nb$_2$FeC \cite{li2021near}. With respect to the ZIA phases, the allowable combinations of elemental constituents could probably be associated with differences in the electronegativity and atomic radius values of the Z-, I- and A-elements; however, before definitive rules governing the ZIA phase formation can be proposed, further systematic research is needed to better map – in terms of attainable chemical compositions – the ZIA phases `landscape'.

Even though the main objective of this work was to introduce the class of nanolaminated ZIA phases via the experimental synthesis and characterization of its Nb$_3$SiNi$_2$ first candidate member as well as to demonstrate the feasibility of producing phase-pure materials based on the Nb$_3$SiNi$_2$ ternary IMC, an overarching stoichiometric rule governing both the MAX and the ZIA phases was discovered and is proposed here. As pointed out by Goossens et al. in a rather recent review \cite{goossens2021microstructure}, the nanolaminated MAX phases are mainly encountered either as the three well-known M$_{n+1}$AX$_n$ phase orders 211, 312 and 413, or as the following exceptional structures: (i) the M$_{n+1}$AX$_n$-like structures with n $>$ 3, such as Ta$_6$AlC$_5$ and Ti$_7$SnC$_6$, and (ii) the ``hybrid'' orders M$_5$A$_2$C$_3$ and M$_7$A$_2$C$_5$, such as Ti$_5$A$_2$C$_3$ with A = Si, Ge and Al. To date, the ``hybrid'' orders M$_5$A$_2$C$_3$ and M$_7$A$_2$C$_5$ have been considered as combinations of a 312 atomic stacking with a 211 and a 413 order, respectively, as they did not obey the generic M$_{n+1}$AX$_n$ phase stoichiometry. However, the careful consideration of both the 312 ZIA phases and all the MAX phase variants, including the ``hybrid'' orders and the only identified member of the 514 order, i.e., \textit{i.e.}, the Mo$_4$VAlC$_4$ compound \cite{deysher2019synthesis}, revealed that both classes of nanolaminated crystalline solids with \textit{fcc} and \textit{hcp} close-packed lattices can be described by the general stoichiometric rule P$_{x+y}$A$_{x}$N$_{y}$ (PAN), where \textit{x} = 1 or 2, and \textit{y} = 1, 2, 3, 4, 5, or 6, according to all the `PAN phases' that were experimentally synthesized to date. The herein proposed global stoichiometric rule governs the constitution of all nanolaminated crystalline solids with atomically close-packed lattices, such as the \textit{fcc} ZIA phases, the \textit{hcp} MAX phases and the ternary \textit{hcp} IMCs, such as Co$_3$SiNb$_2$, Ni$_3$SiNb$_2$ and Cu$_3$SiMg$_2$, even though crystal structural nanolamination has still not been confirmed for the latter. Moreover, this global stoichiometric rule reconciles the ``hybrid'' orders M$_5$A$_2$C$_3$ and M$_7$A$_2$C$_5$ with the common MAX (M$_{n+1}$AX$_n$) phase orders 211, 312 and 413, providing an affirmative answer to the question whether it was conceptually correct to include all identified MAX phase variants in the same class of materials.

A major challenge addressed in this work was to demonstrate that the synthesis of phase-pure bulk ZIA phase-based materials is feasible. Producing phase-pure bulk materials is a prerequisite for the systematic investigation of material properties (\textit{e.g.}, elastoplastic and fracture behavior; oxidation and corrosion resistance; electrical and magnetic properties; radiation tolerance; etc.) or the possible applications of the ZIA phases and their 2D derivatives, provided that these can be extracted. Even though the production of phase-pure Ni$_3$SiNb$_2$ via arc melting proved extremely challenging, presumably due to the volatilization of Ni and/or its compounds above a certain temperature, the synthesis of quasi-phase-pure bulk Ni$_3$SiNb$_2$ via RHP was successful. Similar challenges in producing phase-pure bulk materials have been previously reported in association with the synthesis of specific MAX phases, such as the Zr- and Hf-based MAX phases \cite{lapauw2016synthesis,lapauw2016synthesisA,lapauw2016synthesisB,lapauw2019interaction,tunca2019synthesis,tunca2020compatibility}; typically, overcoming these challenges involved the synthesis of complex double solid solutions based on steric stability criteria \cite{lapauw2018double,tunca2020compatibility}, or the use of early transition metal hydride raw powders rather than elemental ones \cite{goossens2022synthesis}. Apart from RHP, several other processing methods will be explored to synthesize phase-pure ZIA phases in various forms (bulk, coatings) so as to make progress with structural characterization, property determination and possible applications. Physical vapor deposition (PVD) coating techniques, such as magnetron sputtering, are expected to be able to deposit coatings of very precise compositions, such as the herein studied Ni$_3$SiNb$_2$ ZIA phase onto substrates of choice, suppressing the formation of competing phases and producing phase-pure coatings with strong textures that might be desired by certain high-tech applications.

\section{Conclusions}
\label{sec:conclusions}

\noindent Nanostructured materials have revolutionized modern materials science, even though their potential is far from being fully exploited for the benefit of the Society. Nanolaminated ceramics, such as the MAX phases and their 2D derivatives known as MXenes, already changed the way we perceive materials design and functionalization on the nanoscale, thereby producing innovative materials that are characterized by a wide variety of unique properties, precisely tailored to the requirements of the targeted applications. Identifying market niches that could use nanostructured materials, including applications in extreme service environments (\textit{e.g.}, high temperatures and pressures, corrosive media, irradiation, etc.), is an important driver towards the commercial deployment of innovative materials with sophisticated nanodesigns and carefully engineered functionalities. In this work, we propose a novel class of nanolaminated crystalline solids, the `zigzag intermetallic' phases or, simply, the ZIA phases. The herein introduced ZIA phases are characterized by an \textit{fcc} ($Fd\bar{3}m$, space group 227) close-packed lattice of appreciable structural complexity, while obeying the stoichiometric rule of the 312 MAX phases, \textit{i.e.}, M$_{n+1}$AX$_n$ with \textit{n} = 2. The ZIA phases and all known MAX phase orders, including the ``hybrid'' M$_5$A$_2$C$_3$ and M$_7$A$_2$C$_5$ ones mentioned above, appear to respect the herein proposed global stoichiometric rule P$_{x+y}$A$_{x}$N$_{y}$, where \textit{x} and \textit{y} are integers in the 1-6 range. 

The first candidate ZIA phase, \textit{i.e.}, the Nb$_{3}$SiNi$_{2}$ IMC, is identified as the H-phase of the Nb--Si--Ni system and has been found in bulk samples fabricated by both arc melting (low Nb$_{3}$SiNi$_{2}$ content) and RHP (quasi-phase-pure Nb$_{3}$SiNi$_{2}$). The STEM/SAED characterization of Nb$_{3}$SiNi$_{2}$ confirmed the expected diamond cubic crystal structure, revealing a remarkable structural complexity, as reflected in an exceptionally large unit cell made of 96 atoms with nanolayered arrangement. Considering that, apart from the Nb$_{3}$SiNi$_{2}$ IMC, other ternary 312 IMCs (\textit{i.e.}, Mn$_{3}$SiNi$_{2}$, V$_{3}$SiNi$_{2}$, Cr$_{3}$SiNi$_{2}$, Ta$_{3}$SiNi$_{2}$, Nb$_{3}$SiCo$_{2}$, Na$_{3}$AuIn$_{2}$, Na$_{3}$AgIn$_{2}$, and Mg$_{3}$GaNi$_{2}$) with the same diamond cubic \textit{fcc} lattice have already been experimentally synthesized, it is reasonable to assume that these IMCs constitute an entire class of nanolaminated crystalline solids, herein designated as the ZIA phases. Confirming the nanolaminated nature of the other ternary 312 IMCs is the aim of ongoing investigations. 

The ZIA phases extend the concept of crystal structure nanolamination beyond the early transition metal carbides/nitrides known as the MAX phases. Despite sharing similarities and the same historical roots with the MAX phases, ZIA phases such as the Nb$_{3}$SiNi$_{2}$ IMC are characterized by appreciably more complex crystal structures and, by extension, configurational entropies than ternary MAX phases such as the Ti$_{3}$SiC$_{2}$ ceramic. Therefore, the nanolaminated ZIA phases are conceptually positioned, in terms of characteristics and potential applications, between the nanolaminated MAX phases (with unique hybrid metallic/ceramic properties, and proven capability of producing 2D derivatives, known as MXenes, with tailored functionalities) and other dense and complex multicomponent alloys, such as high-entropy alloys (HEAs) characterized by large chemical complexity and remarkable radiation tolerance. This work succeeded in synthesizing phase-pure bulk Nb$_{3}$SiNi$_{2}$ via reactive hot pressing (RHP), which is considered an important achievement as the production of phase-pure ZIA phase-based materials is not only a prerequisite for the (mechanical, electrical, magnetic, etc.) property determination of these largely unknown compounds, but also to explore the possibility of producing and characterizing their (potential) 2D derivatives, as well as assessing ways to tailor their properties (\textit{e.g.}, ductility) in an application-driven manner. It is likely that the new class of ZIA phases will create a new ecosystem of nanolaminated/nanostructured (3D/2D) materials with amazing new properties and fascinating applications in diverse technological areas. The first step towards the exploration of this hitherto unexplored ecosystem is the successful synthesis of many materials with high phase purity, starting from the already identified candidate ZIA phases (\textit{e.g.}, Nb$_{3}$SiNi$_{2}$, Mn$_{3}$SiNi$_{2}$, V$_{3}$SiNi$_{2}$, Cr$_{3}$SiNi$_{2}$, Na$_3$AuIn$_2$, Mg$_3$GaNi$_2$, etc.) and proceeding with theoretically predicted (\textit{e.g.}, by means of thermodynamic modeling, density functional theory, etc.) but not yet experimentally made IMCs. The second step in this exploratory work should include attempts to produce 2D derivatives of the ZIA phases, as well as ternary \textit{hcp} IMCs (\textit{e.g.}, Co$_3$SiNb$_2$, Ni$_3$SiNb$_2$, Cu$_3$SiMg$_2$, etc.), in order to find out whether they are also characterized by crystal structure nanolamination, as expected, or not.

\section{Experimental Section}
\label{sec:methods}

\subsection{Synthesis via Arc Melting}
\label{sec:methods:synthesis_arc}

\noindent Conventional vacuum arc melting of high-purity ($>$99.99\%) Nb, Ni and Si elemental raw powders was initially employed to synthesize the `Nb–Si–Ni intermetallic alloy' with the targeted H-phase Nb$_{3}$SiNi$_{2}$ stoichiometry, \textit{i.e.}, 50.0Nb-16.7Si-33.3Ni\footnote{The term ``Nb–Si–Ni intermetallic alloy'' is used for this alloy composition throughout the manuscript.}, in at.\%. A disc (200 g) of low phase purity was produced via arc melting, despite the fact that the ratio of elemental powders in the powder feedstock had been adjusted to the Nb$_{3}$SiNi$_{2}$ stoichiometry. Additional information on the arc melting synthesis of alloys in the Nb--Si--Ni system may be found elsewhere \cite{vinicius2015thermodynamic,vinicius2019experimental,vinicius2020experimental}.

The thermal and microstructural stability of the as-cast Nb–Si–Ni intermetallic alloy was assessed via an isothermal annealing treatment at 1421 K for 2 weeks (336 h) in a vacuum furnace (ThermoLyne furnace, model 46100). The as-cast alloy samples were pre-encapsulated in a quartz (SiO$_2$) tube, which was pressurized with argon (Ar) gas to 1 atm; both heating and cooling rates were fixed at about 14.2 K/min.

\subsection{Synthesis via Reactive Hot Pressing}
\label{sec:methods:synthesis_RHP}

\noindent The inability to produce a phase-pure Nb–Si–Ni intermetallic alloy by means of arc melting redirected the synthesis efforts in this work towards a powder metallurgical processing route. Quasi phase-pure discs of the Nb–Si–Ni intermetallic alloy were successfully produced via reactive hot pressing (RHP), using hydride and elemental starting powders, \textit{i.e.}, NbH$_{0.89}$ ($<$40 $\mu$m, CBMM, Brazil), Ni (Vale, T123\texttrademark, 3-7 $\mu$m, UK) and Si (A10, 2.1 $\mu$m, HC Starck, Germany). To optimize the mixing of the starting powders and improve powder particle contact, the as-received NbH$_{0.89}$ powder was first refined by low-energy ball milling in isopropanol for 24 h (Turbula\textregistered T2), using WC-6 wt.\% Co (WC-6Co) milling balls (diameters of 5 and 10 mm). The refined NbH$_{0.89}$ powder was dried in a rotating evaporator (Heidolph 4010) and then mixed with the Ni and Si elemental powders in isopropanol for another 24 h, using WC-6Co balls. The ratio of the NbH$_{0.89}$, Ni and Si powders in the powder feedstock was adjusted to the targeted 3Nb:1Si:2Ni. After mixing, the powder suspension was dried in a rotating evaporator.

The powder feedstock was cold-pressed into `green' pellets (30 mm in diameter, 4-5 mm thickness) under a load of 30 MPa in a graphite die. Exploratory RHP runs were carried out in a hot press (W100/150-2200-50 LAX, FCT Systeme, Frankenblick, Germany). The powder compacts were heated in vacuum (0.4 mbar) at 10 K/min to different sintering temperatures (\textit{i.e.}, 1523, 1623 and 1723 K), where consolidation occurred under an uniaxial pressure of 30 MPa for a dwell period of 60 min. The sintered discs (30 mm in diameter) were allowed to cool naturally by switching off the power supply of the hot-press.

\subsection{Materials Characterization}
\label{sec:methods:characterization}

\noindent Metallographic cross-sections were prepared from all Nb–Si–Ni intermetallic alloy samples produced via arc melting and RHP; all cross-sections were polished to a mirror surface finish with a colloidal silica (OPS) suspension in the final step. The microstructure of as-cast and isothermally annealed samples produced via arc melting was studied by means of a Thermo Fisher Apreo SEM (scanning electron microscope) equipped with an EDX (energy-dispersive X-ray spectroscopy) detector (EDAX Octane Elite Super). Elemental mapping in the RHP samples was done by means of an electron probe microanalyzer (EPMA; JXA-8530F, JEOL Ltd, Japan) equipped with a WDXS (wavelength-dispersive X-ray spectroscopy) detector (1-5 eV, 8 eV at Fe-K$_{\alpha}$, full scanner type spectrometer). EDX elemental analysis (1 SDD/UTW, 129 eV at Mn-K$_{\alpha}$) of the identified phases was performed on the same device at fixed conditions (15 kV accelerating voltage, 100 nA beam current).

X-ray diffraction (XRD) was employed to identify the phases present in all Nb–Si–Ni intermetallic alloy samples at room temperature. The as-cast and annealed samples produced via arc melting were analyzed on a Bruker AXS D8 Advance X-ray diffractometer, using Cu-K$_{\alpha}$ radiation (40 kV, 40 mA) in a Bragg-Brentano geometry; the XRD spectra were recorded at steps of 0.02$^{\circ}$ in the 20-90$^{\circ}$ 2$\theta$ range. The RHP samples were analyzed on a Rigaku SmartLab SE XRD system with 1D/teX Ultra 250 HE detector, using Cu-K$_{\alpha}$ radiation (40 kV, 40 mA) in a Bragg-Brentano geometry; the XRD spectra were recorded at steps of 0.05$^{\circ}$ in the 5-90$^{\circ}$ 2$\theta$ range, at a scan speed of 0.1$^{\circ}$ per min.

An FEI Helios 600 dual-beam FIB/SEM was used to lift out TEM (transmission electron microscopy) thin foils from sufficiently large Nb$_{3}$SiNi$_{2}$ grains in the arc-melted alloy samples, utilizing the classic lift-out method \cite{giannuzzi1999review}. Combined Pt/C protective capping layers were deposited on the sites of interest to protect the foils from ion beam damage during lift-out. The FIB (focused ion beam) foils were ion-milled to electron transparency at 30 keV, with final thinning/ion beam damage removal at 2 keV.

TEM analysis of the FIB foils lifted out from the arc-melted alloy samples was performed by means of a FEI Tecnai TF30 and a FEI Titan 80/300, both operated at 300 kV. The latter was equipped with an EDAX Octane Elite T EDX detector, and all EDX quantification work was conducted based on the Cliff-Lorimer method \cite{cliff1972quantitative,cliff1975quantitative}. Selected area electron diffraction patterns (SAEDPs) of the Nb$_{3}$SiNi$_{2}$ IMC ZIA phase were indexed using data from the Inorganic Crystal Structure Database (ICSD) \cite{gladyshevskii1964crystal}. The CrystalMaker, SingleCrystal and CrystalDiffraction software packages were used to simulate electron diffraction patterns and aid the indexing of experimentally acquired SAEDPs \cite{palmer2015visualization}.

\subsection{Thermodynamic \& Crystal Structure Simulations}
\label{sec:methods:calphad}

\noindent Thermodynamic simulations were performed to predict the equilibrium phases at the sintering temperatures of the RHP alloys (\textit{i.e.}, 1523 K, 1623 K, and 1723 K), as well as at 1423 K (\textit{i.e.}, close to the homogenization annealing temperature, 1421 K, of the arc-melted alloy). These simulations were performed with version V2020a of the Thermo-Calc software (Thermo-Calc Software Inc., Solna, Sweden) \cite{sundman1985thermo}, based on the database published by dos Santos \textit{et al.} \cite{vinicius2015thermodynamic}. The unit cells of the \textit{fcc} Nb$_{3}$SiNi$_{2}$  ZIA phase and the \textit{hcp} Ti$_{3}$SiC$_{2}$ MAX phase were drawn using version 3.5.8 of the VESTA software \cite{momma2011vesta}, utilizing the Wyckoff parameters provided in Table S1 (Supplementary Information) for precise crystal structure representation. These two phases were compared in terms of relative structural complexity and unit cell sizes, whilst certain zone axes of the herein addressed Nb$_{3}$SiNi$_{2}$ ZIA phase were selected to depict `zigzag' atomic arrangement and crystal structure nanolamination.

\section*{Acknowledgments}
\label{sec:acknownledgements}

\noindent The Los Alamos National Laboratory (LANL), an affirmative action equal opportunity employer, is managed by Triad National Security, LLC, for the U.S. Department of Energy’s National Nuclear Security Administration, under contract number 89233218CNA000001. LANL provided research support to MAT via the Laboratory Directed Research and Development (LDRD) program, under project number 20200689PRD2. OEA acknowledges funding from his Early Career Program supported by LANL's LDRD, under contract number 20210626ECR. CGS thanks the financial support provided by the Conselho Nacional de Desenvolvimento Científico e Tecnológico (CNPq, Brasília-DF, Project 307627/2021-7). MAT would like to thank Dr. Kurt E. Sickafus, LANL, for valuable discussions on the crystallography of complex solid-state phases. KL and JV acknowledge funding from the Euratom research and training program 2014-2018 under Grant Agreement No. 740415 (H2020 IL TROVATORE). NG acknowledges the Fund for Scientific Research Flanders (FWO-Vlaanderen) for funding his PhD Fundamental Research Fellowship No 1118120N.

\section*{Author Contributions}
\label{sec:contributions}

\noindent The initially devised experimental plan that employed arc melting to explore the precursor idea of the ZIA phases was formulated by MAT as a project for a Director's fellowship at LANL. MAT and KL share the conceptualization of the major scientific ideas associated with the class of nanolaminated ternary IMCs herein identified as the ZIA phases as well as their correlation with the class of nanolaminated early transition metal carbides/nitrides known as the MAX phases. SH, NG, and JV succeeded in producing quasi-phase-pure Nb$_{3}$SiNi$_{2}$ materials by means of RHP. All other authors have smaller contributions with experiments, data analysis, research supervision, and funding. All authors have read and approved the manuscript. MAT started the first draft and KL extensively revised it, contributing with key ideas and RHP data, the latter in collaboration with KU Leuven.
\newpage

\section*{References}
\label{sec:references}
\bibliography{bibdata}
\bibliographystyle{elsarticle-num}

\end{document}